\documentclass[preprintnumbers, prd, showpacs, floatfix, preprintnumbers, letterpaper, nofootinbib, amsmath, amssymb, superscriptaddress,preprint]
{revtex4-1}
\usepackage[T1]{fontenc}
\usepackage{graphicx}
\usepackage{epstopdf}
\usepackage{epsfig}
\usepackage{bm}
\usepackage{amsfonts}

\usepackage[latin1]{inputenc}
\usepackage{graphicx}
\usepackage[english]{babel}
\usepackage{amsmath}
\usepackage{amssymb}

\usepackage{color}

\newbox\pippobox

\def\be{\begin{equation}}
\def\ee{\end{equation}}
\def\ba{\begin{eqnarray} }
\def\ea{\end{eqnarray}}

\newcommand {\lla} {\ {\raise-.5ex\hbox{$\buildrel<\over\sim$}}\ }
\renewcommand{\(}{\left(}
\renewcommand{\)}{\right)}
\renewcommand{\[}{\left[}
\renewcommand{\]}{\right]}

\begin{document}

\title{Cosmological model due to dimensional reduction of higher-dimensional massive gravity theory}
\author{Ratchaphat Nakarachinda \footnote{Email: tahpahctar\_net@hotmail.com}}
\affiliation{The institute for fundamental study, Naresuan University, Phitsanulok 65000, Thailand}

\author{Pitayuth Wongjun \footnote{Email: pitbaa@gmail.com}}
\affiliation{The institute for fundamental study, Naresuan University, Phitsanulok 65000, Thailand}
\affiliation{Thailand Center of Excellence in Physics, Ministry of Education,
Bangkok 10400, Thailand}

\begin{abstract}
We investigate a cosmological model resulting from a dimensional reduction of the higher-dimensional dRGT massive gravity. By using the Kaluza-Klein dimensional reduction, we obtain an effective four-dimensional massive gravity theory with a scalar field. It is found that the resulting theory corresponds to a combined description of mass-varying massive gravity and quasi-dilaton massive gravity. By analyzing the cosmological solution, we found that it is possible to obtain the late-time expansion of the universe due to the graviton mass. By using a dynamical system approach, we found regions of model parameters for which the late-time expansion of the universe is a stable fixed point. Moreover, this also provides a mechanism to stabilize the extra dimensions. 

\end{abstract}
\maketitle

\section{Introduction}

It is well-known that the universe is expanding with acceleration nowadays \cite{Riess:1998cb,Perlmutter:1998np}. By using general relativity (GR) with cosmological constant, the theoretical predictions satisfy most of observational results. However, the value of the cosmological constant needs to be fine-tuned to the order of $10^{-122}$ in Planck units. This does not provide a proper model in theoretical point of view. As a result, alternatively theoretical models have been intensively constructed in order to describe the accelerating expansion of the universe. One of possible ways to construct this theoretical model is to modify GR at large scale, driving the expansion at cosmological scale while recovering GR at local gravity scale. In this work, we focus on the modification of GR by adding  suitable mass term into Einstein-Hilbert action proposed by de Rham, Gabadadze and Tolley \cite{deRham:2010ik, deRham:2010kj} namely dRGT massive gravity theory.

In dRGT massive gravity theory, the mass terms are systematically constructed in order to eliminated the Boulware-Deser (BD) ghost from the theory. The BD ghost is an additional sixth degree of freedom which usually emerges in nonlinear massive gravity theory \cite{Boulware:1973my}. This healthy nonlinear massive gravity then provides a new window to various researches, not only in cosmological context but also in astronomical objects and black holes \cite{Koyama:2011yg,Koyama:2011xz,Cai:2014znn, Ghosh:2015cva, Boonserm:2017qcq}. In cosmological context, it was found that the theory with Minkowski fiducial metric does not admit flat and closed Friedmann-La\^irmatre-Robertson-Walker (FLRW) solutions \cite{Gumrukcuoglu:2011ew}. In order to obtain all kinds of FLRW solutions, it has been suggested to consider three options as follows \cite{D'Amico:2011jj}: 1) Use other forms of fiducial metric, for example, FLRW  or de-Sitter \cite{Gumrukcuoglu:2011zh,Fasiello:2012rw, Langlois:2012hk, Langlois:2013cya}, 2) Consider anisotropic universe \cite{DeFelice:2012ac,Mukohyama:2012op, Antonio:2013qr} and 3) Add more degree of freedom into the theory. Although the self-accelerating expansion of the universe can be provided by using the FLRW fiducial metric, there are only two propagating degrees of freedom while there should be five for massive gravity theory in four-dimensional spacetime \cite{Gumrukcuoglu:2011zh}. Adding more degrees of freedom is still one of possible ways for solving the lack of the number of propagating degrees of freedom. In this work, we will use the flat FLRW fiducial metric and specialize our attention to the third option where the additional degree of freedom is a scalar field.

The dRGT massive gravity models with additional degree of freedom have been investigated in various ways. By promoting the fiducial metric to be dynamical, the theory is known as "bi-gravity theory" \cite{Hassan:2011zd}. Other class of this extension is obtained by adding a scalar field into the theory. One of simple arguments is that promoting the graviton mass to be varied according to a scalar field called mass-varying massive gravity (MVMG) \cite{Huang:2012pe,Saridakis:2012jy, Wu:2013ii,Leon:2013qh,Huang:2013mha}. One of drawback features of this model is that the accelerating expansion of the universe is driven by the scalar field instead of the graviton mass, the graviton mass shrinks to zero at late time expansion of the universe. Fortunately, the graviton mass can play a role of the cosmological constant by using the FLRW form of the fiducial metric \cite{Tannukij:2015wmn}. Further investigation also shows that a part of graviton mass can be responsible for dark matter by promoting the scalar field as the k-essence field \cite{Tannukij:2015wmn}.

Another extension of dRGT massive gravity by using scalar field is quasi-dilaton massive gravity (QMG) \cite{D'Amico:2012zv,Gannouji:2013rwa}. The additional scalar field possesses the quasi-dilaton global symmetry. It was found that the self-accelerating solution to the model plagued by ghost instability \cite{Gumrukcuoglu:2013nza, DAmico:2013saf, Haghani:2013eya}. Further extended models of the QMG are investigated  by introducing the coupling between the scalar field and the Stuckelberg fields  \cite{DeFelice:2013tsa, DeFelice:2013dua} and then found that it is possible to avoid the ghost instability in some classes of the solutions \cite{Mukohyama:2013raa,Kluson:2013jea,Mukohyama:2014rca, Heisenberg:2015voa, DeFelice:2016tiu, DeFelice:2017wel, Anselmi:2017hwr, Gumrukcuoglu:2017ioy, DeFelice:2017rli}. Another extension of the theory is investigated by  introducing the Dirac-Born-Infeld (DBI) scalar field in such a way that the scalar field possesses the generalized Galileon shift symmetry \cite{Hinterbichler:2013dv, Gabadadze:2012tr, Andrews:2013uca}. Unfortunately, the model encounters problems such that the number degree of freedom is not correct or the ghost mode emerges in the linear perturbation level \cite{Chullaphan:2015ija}.

The theories with higher-dimensional spacetime have been intensively motivated from a more fundamental theory such as String theory (or M theory). Since the observations suggest that we live in four-dimensional spacetime, a mechanism to reduce the higher-dimensional spacetime to four-dimensional spacetime must be involved and one of well-known mechanisms is the Kaluza-Klein (KK) dimensional reduction \cite{Kaluza:1921tu,Klein:1926tv}. Since gravity theories in higher-dimensional spacetime have more degree of freedom than ones in four-dimensional spacetime, it is generally to obtain the four-dimensional gravity theory with some additional fields associated with the structure of the extra dimensions. It is important to note that cosmological aspect of dRGT massive gravity with higher-dimensional spacetime has been investigated in the context of braneworld scenario \cite{Yamashita:2014cra} where the matter fields are confined in three-dimensional hypersurface \cite{Randall:1999ee,Randall:1999vf}. In this work, we focus on other kinds of the higher-dimensional gravity theory where the matter fields can access to all dimensions called Universal Extra Dimension (UED) \cite{Appelquist:2000nn}. 
Note also that besides a dRGT theory, a dimensional reduction of a partially massless theory, which is closely related to a linear massive gravity, has been investigated \cite{Bonifacio:2016blz}. Due to the simple structure of the extra dimensions, the additional field is a scalar field characterizing the radius of the extra dimensions. Therefore, it is worthwhile to consider the dRGT massive gravity theory in higher-dimensional spacetime and choose a simple ansatz, assuming the maximally symmetric extra dimensions, to obtain a four-dimensional dRGT massive gravity theory with an additional scalar field. According to this argument, we propose an alternative extension of dRGT massive gravity by using a scalar field which is obtained from the scalar degree of freedom characterizing the radius of the volume of the extra dimensions and then find a possibility to provide a cosmological solution such that the late-time expansion of the universe is obtained.

This paper is organized as follows: in Section \ref{sec:HDMG and reduction}, we consider dRGT massive gravity theory in (4+d)-dimensional spacetime and then perform the KK dimensional reduction. As a result, we obtain an effective four-dimensional massive gravity theory with a scalar field. It is found that the theory admits a combined description of MVMG and QMG.  In Section \ref{sec:cosmology}, we investigate a cosmological model of this kind of massive gravity theory and found that it is possible to obtain the late-time expansion of the universe. In Section \ref{sec:dynamical system}, we investigate this possibility by using the dynamical system approach to analyze dynamics of the universe. We found some regions of the model parameters that provide the standard evolution of the universe as well as the extra dimension can be stabilized. Finally, we summarize and discuss the results in Section \ref{sec:conclusion}.


\section{Dimensional reduction of ($4+d$)-dimensional dRGT massive gravity theory}\label{sec:HDMG and reduction}
One of possible ways to construct a consistent theory of massive gravity is to consider the Einstein gravity theory in ($4+d$)-dimensional spacetime where $d$ is the number of the extra dimensions. In fact it was shown that the dRGT massive gravity theory can be obtained by using discretization procedure of pure massless gravity in ($4+d$)-dimensional spacetime 
\cite{Hinterbichler:2012cn,Deffayet:2012zc,deRham:2014zqa}. As we have mentioned, one of possible ways to obtain the cosmological solution in dRGT massive gravity is adding the additional field into the theory. Since the scalar degree of freedom can be obtained from dimensional reduction, it is worthwhile to consider a reduced theory of the ($4+d$)-dimensional dRGT massive gravity theory. In this section, we will perform a dimensional reduction of the ($4+d$)-dimensional dRGT massive gravity theory. Let us consider the dRGT massive gravity theory in ($4+d$)-dimensional spacetime as follows \cite{Do:2016abo}
\begin{eqnarray}\label{4+d action}
	S=\int \text{d}^{4+d}x\sqrt{-\tilde{g}}\frac{M^{d+2}_{(4+d)}}{2}\left(\tilde{R}+m^2_g\tilde{U}\right),\,\,\,\,\,\,\,\,
	{\tilde U} = {\tilde U}_2+\sum_{j=3}^{4+d}\alpha_j{\tilde U}_j,
\end{eqnarray}
where quantities with tilde, $\tilde{X}$, refer to the quantities in ($4+d$)-dimensional spacetime. $\tilde{R}$ is the Ricci scalar, $M_{(4+d)}$ is the  Planck mass in ($4+d$)-dimensional spacetime, $m_g$ is the parameter associated to the graviton mass. Each potential ${\tilde U}_i$ is a function of $[\tilde{K}^n]$ where ${\tilde K}^A_B=\delta^A_B-\sqrt{{\tilde g}^{-1}{\tilde f}\,}^A_B$ and $[{\tilde K}^n]=\text{tr}(({\tilde K}^n)^A_B)$. The explicit form of $\tilde{U}_i$ is presented in Appendix \ref{app:Ut}. The metric ${\tilde f}_{AB}$ is the reference or fiducial metric and capital Latin indices refer to ($4+d$)-dimensional spacetime indices. In order to perform the KK dimensional reduction which keeps only the scalar field left, we neglect the cross components between the extra dimension and the four-dimensional spacetime. As a result, the physical metric can be written as
\begin{eqnarray}
	{\tilde g}_{AB} =\(\begin{array}{cc}p^{-d}(\phi)g_{\mu\nu}& 0\\0&p^2(\phi)\gamma_{ab}\end{array}\),
\end{eqnarray}
where $g_{\mu\nu}$ is the metric in four-dimensional spacetime, $\gamma_{ab}$ is the metric in the extra dimensions, the small Latin indices refer to the indices in extra dimensions, and $p(\phi)$ is a function playing a role of the radius of the extra dimensions. In the same way as the physical metric, the fiducial metric in ($4+d$)-dimensional spacetime can be decomposed as
\begin{eqnarray}
	{\tilde f}_{AB} =\(\begin{array}{cc}f_{\mu\nu}&0\\0&q^2(\phi)\gamma_{ab}\end{array}\),
\end{eqnarray}
where $f_{\mu\nu}$ is the fiducial metric in four-dimensional spacetime and $q(\phi)$ is a function playing the role of the radius of the extra dimensions in the fiducial sector. For this ansatz, the action in Eq. \eqref{4+d action} can be rewritten as
\begin{eqnarray}\label{extend 4+d action}
	S&=&\int \text{d}^4x \text{d}^dyp^{-d}\sqrt{-g}\sqrt{\gamma}\frac{M^{d+2}_{(4+d)}}{2}\[\begin{array}{l}p^d\left\{\begin{array}{l}
	R\[g\]+p^{-(d+2)}R\[\gamma\]-\frac{d(d+2)}{2}\frac{p^2_{,\phi}}{p^2}\nabla_\rho\phi\nabla^\rho\phi\\
	+2(d-1)\nabla_\rho\(\frac{1}{p}\nabla^\rho p\)
	\end{array}\right\}\\+m^2_g\(U+d\,rF\)\end{array}\],
\end{eqnarray}	
where $y^a$ are coordinates on the extra-dimensional manifold which is assumed to be maximally symmetric  with the metric $\gamma_{ab}$. $R[\gamma]$ is the Ricci scalar associated with the metric $\gamma_{ab}$. Since we deal with the maximally symmetric (extra) space, the curvature parameter of $\gamma_{ab}$ for $d>1$ is expressed as $\kappa=\frac{R[\gamma]}{d(d-1)}$. Note that this curvature parameter for $d=1$ is undefined and the geometry of the extra dimensions is hyperbolic, flat and spherical for $\kappa <0$, $\kappa = 0$ and $\kappa > 0$ respectively. 
The potential in the mass term $\tilde{U}$ can also be splitted into the usual form of the four-dimensional one, $U$, and the additional term, $F$. The usual potential, $U$ is taken the same form as $\tilde{U}$ but put $[K^n]$ instead of $[\tilde{K}^n]$ where 
\begin{eqnarray}
	K^\mu_\nu=\delta^\mu_\nu-p^{d/2}\sqrt{g^{-1}f\,}^\mu_\nu.
\end{eqnarray}
For the additional potential, $ F$, can be written as  $F=F_2+\sum_{j=3}^{4+d}\alpha_jF_j$ where $r=1-q/p$. Each $F_i$ is written explicitly in Appendix \ref{app:F}. Now we will perform the dimensional reduction of the ($4+d$)-dimensional action in Eq. \eqref{extend 4+d action}. As a result, the effective four-dimensional action becomes
\begin{eqnarray}\label{4 action}
	S&=&\int \text{d}^4x\sqrt{-g}\frac{m^2_{pl}}{2}\left\{R\[g\]-\frac{2}{m^2_{pl}}V(\phi)-\frac{1}{m^2_{pl}}\nabla_\rho\phi\nabla^\rho\phi+M^2_g\(U+d\,rF\)\right\},
\end{eqnarray}
where the four-dimensional Planck mass, $m_{pl}$, is related to the ($4+d$)-dimensional Planck mass, $M_{(4+d)}$, via $m^2_{pl}=\int \text{d}^dy \sqrt{\gamma} M^{d+2}_{(4+d)}$. The potential of the scalar field, $V(\phi)$ is defined by $V(\phi)\equiv-\frac{1}{2}d(d-1)m^2_{pl}p^{-(d+2)}\kappa$. From this scalar potential, one can see that it vanishes for the case of five-dimensional spacetime or if the spacetime geometry of the extra dimension is flat.  The canonical form of the scalar field can be obtained by setting $p=e^{\sqrt{\frac{2}{d(d+2)}}\frac{\phi}{m_{pl}}}$. As we have seen, the graviton mass, $M_g$ is promoted as a function of the scalar field,
\begin{eqnarray}
	M^2_g(\phi)\equiv p^{-d}m^2_g=e^{-\sqrt{\frac{2d}{d+2}}\frac{\phi}{m_{pl}}}m^2_g.
\end{eqnarray}
Note that the radius function of the fiducial metric, $q(\phi)$, is arbitrary. For the simplest case, one can choose as the same form as the radius function of the physical metric. This corresponds to choosing $r=0$ so that the potential reduces to one in four-dimensional massive gravity. Moreover, the potential $U$ also respects a global symmetry
\begin{eqnarray}
	\phi \rightarrow \phi + \phi_0,\,\,\,\,\, \psi^a \rightarrow e^{-\sqrt{\frac{d}{2(d+2)}}\frac{\phi_0}{m_{pl}}} \psi^a,
\end{eqnarray}
where $\psi^a$ are four Stuckelberg fields defined via the fiducial metric as $f_{\mu\nu} \equiv f_{ab} \partial_\mu\psi^a \partial_\nu\psi^b$. Therefore, the structure of the potential is exactly the same with QMG
\cite{D'Amico:2012zv,Gannouji:2013rwa}. However, the mass function, $M_g(\phi)$, breaks this global symmetry. Since the mass function depends on the scalar field, the theory is equivalent to MVMG \cite{Huang:2012pe,Saridakis:2012jy, Wu:2013ii,Leon:2013qh,Huang:2013mha}. As a result, the theory is somewhat a combination of QMG and MVMG. Up to our knowledge, this kind of extended dRGT massive gravity has not been investigated yet. Therefore, the main purpose of this paper is to investigate the cosmological models due to this kind of massive gravity.

Since the theory contains two dynamical fields, $g_{\mu\nu}$ and $\phi$, equations of motion can be obtained by varying the action in Eq. \eqref{4 action} with respect to both fields. As a result, the equations of motion for $g_{\mu\nu}$ can be expressed as
\begin{eqnarray}\label{eom for guv}
	 G_{\mu\nu}-\frac{1}{m^2_{pl}}g_{\mu\nu}V+\frac{1}{2m^2_{pl}}g_{\mu\nu}\nabla_\rho\phi\nabla^\rho\phi-\frac{1}{m^2_{pl}}\nabla_\mu\phi\nabla_\nu\phi+M^2_g\(X_{\mu\nu}+d\,rY_{\mu\nu}\)=0
\end{eqnarray}
where $X_{\mu\nu}=\frac{\delta U}{\delta g^{\mu\nu}}-\frac{1}{2}g_{\mu\nu}U$ and $Y_{\mu\nu}=\frac{\delta F}{\delta g^{\mu\nu}}-\frac{1}{2}g_{\mu\nu}F$. The explicit forms are lengthy, they will be presented in Appendix \ref{app:Xuv Yuv}. It is convenient to rewrite Eq. \eqref{eom for guv} in terms of the energy momentum tensors of the scalar field and the mass term as follows
\begin{eqnarray}\label{mo-Ein-eq}
	G_{\mu\nu}=m^{-2}_{pl} \left(T^{(X)}_{\mu\nu}+T^{(\phi)}_{\mu\nu}\right),
\end{eqnarray}
where
\begin{eqnarray}
	T^{(X)}_{\mu\nu}=-m^{2}_{pl}  M^2_g\(X_{\mu\nu}+d\,rY_{\mu\nu}\),\,\,\,\,\,\,\text{and}\,\,\,\,\,\,
	T^{(\phi)}_{\mu\nu}=\nabla_\mu\phi\nabla_\nu\phi-\frac{1}{2}g_{\mu\nu}\nabla_\rho\phi\nabla^\rho\phi-g_{\mu\nu}V.
\end{eqnarray}
By using the Bianchi identity, the conservation of the total energy momentum tensor can be obtained as follows
\begin{eqnarray}\label{const-eq}
	\nabla_\mu\left\{ \(T^{(X)}\)^\mu_\nu+ \(T^{(\phi)}\)^\mu_\nu\right\}= 0.
\end{eqnarray}
Note that since both contents are coupled, they are not separately conserved. Actually, it can be viewed as a coupling model between a scalar field and the content corresponding to the graviton mass. We will show this argument explicitly in the next section. Note also that this conserved equation corresponds to the equation obtained from varying the action with respect to the fiducial metric. This will provide a constraint equation since the fiducial metric plays the role of Lagrange multiplier in the language of Lagrangian formulation.

The equation of motion corresponding to varying the action in Eq. \eqref{4 action} with respect to the scalar field can be written as
\begin{eqnarray}\label{eom for phi}
	\nabla_\rho\nabla^\rho\phi-V_{,\phi}+\sqrt{\frac{d}{2(d+2)}}m_{pl}M^2_g\(\Phi+d\,r\Psi\)=0,
\end{eqnarray}
where $\Phi=m_{pl}\sqrt{\frac{d+2}{2d}}\frac{\delta U}{\delta\phi}-U$ and $\Psi=m_{pl}\sqrt{\frac{d+2}{2d}}\frac{\delta F}{\delta\phi}-F$ are presented in Appendix \ref{app:Phi Psi}.

The field equations, Eq. \eqref{eom for guv} and Eq. \eqref{eom for phi} are all equations of motion in this effective massive gravity theory. Furthermore, these equations of motion are also found from the dimensional reduction at the equations of motion in ($4+d$)-dimensional spacetime; the equations of motion for $g^{\mu\nu}$ in Eq. \eqref{eom for guv}, are obtained from ($\mu,\nu$)-components of the equations of motion for $\tilde{g}^{AB}$ and the equation of motion for $\phi$ in Eq. \eqref{eom for phi}, is obtained from ($a,b$)-components of the equations of motion for $\tilde{g}^{AB}$ combined with the ($a,b$)-components of the constraints derived by varying the action in Eq. \eqref{4+d action} with respect to $\tilde{f}^{AB}$.


\section{Cosmological model}\label{sec:cosmology}

As we have mentioned, the pure dRGT massive gravity does not admit  the flat Friedmann-Lema\^itre-Robertson-Walker (FLRW) solution. Then it may not possibly be a cosmological model for the late-time expansion of the universe. As one of possible ways to obtain the consistent model, the additional degree of freedom can be taken into account. In the previous section, we have shown that the additional scalar field can be obtained by considering the effective theory dimensionally reduced from the dRGT massive gravity theory in ($4+d$)-dimensional spacetime. In this section, we will investigate the possibility to obtain the consistent cosmological solutions for the late-time expansion of the universe from the reduced theory.

In order to investigate the cosmological solutions, let us adopt the physical metric as the flat FLRW metric as follows
\begin{eqnarray}
	\text{d}s^2 = - \text{d}t^2 + a(t)^2 \delta_{ij} \text{d}x^i \text{d}x^j,
\end{eqnarray}
where $a(t)$ is a scale factor determining the scale of the spatial distance. Moreover, it is convenient to work in the unitary gauge where $\psi^a = x^a$ and then choose the form of the fiducial metric as the flat FLRW metric,
\begin{eqnarray}
	f_{\mu\nu} = \text{diag}(-1,b(t)^2,b(t)^2,b(t)^2),
\end{eqnarray}
where $b(t)$ is a scale factor for the spatial distance in fiducial metric. It is instructive and worthwhile to consider the simple case by choosing the radius function as $q=p$, corresponding to $r=0$. This is instructive but useful for a more general case. Including Lagrangian density of the matter $\mathcal{L}_m$ and radiation $\mathcal{L}_r$, the action of the model can be written as
\begin{eqnarray}
	S&=&\int d^4x \sqrt{-g}\left\{\frac{m^2_{pl}}{2}\( R+M_g^2(\phi)U \) -\frac{1}{2} \nabla_\rho\phi \nabla^\rho\phi-g_{\mu\nu}V+\mathcal{L}_m+\mathcal{L}_r\right\},\label{4D-action-matter-rad}
\end{eqnarray}
As a result, the modified Einstein field equation in Eq. \eqref{mo-Ein-eq} becomes
\begin{eqnarray}
	G_{\mu\nu}  = m^{-2}_{pl} \left( T^{(X)}_{\mu\nu}+T^{(\phi)}_{\mu\nu}+T^{(m)}_{\mu\nu}+T^{(r)}_{\mu\nu} \right),\label{mo-Ein-matt-rad-eq}
\end{eqnarray}
By using ansatz we adopt in this section, the components $(0,0)$ and $(i,j)$ can be written respectively as
\begin{eqnarray}
	3 m^{2}_{pl}H^2 &=&m^{2}_{pl}M^2_g A +\(\frac{1}{2}\dot{\phi}^2+V\) +\rho_m+\rho_r, \label{Ein00}\\
m^{2}_{pl}(2\dot{H}+3H^2) &=&m^{2}_{pl}M^2_g B -\(\frac{1}{2}\dot{\phi}^2-V\) - p_m- p_r,\label{Einij}
\end{eqnarray}
where $H$ is the Hubble parameter, $\rho_m$ and $p_m$ are energy density and pressure of the matter respectively as well as $\rho_r$ and $p_r$ are ones of the radiation. The short-hand function $A$ and $B$ obtained from the tensor $X^\mu_\nu$ can be expressed as
\begin{eqnarray}
	A &=& X^0_0  = -3(y+\alpha y^2+\beta y^3), \,\,\,\,\, y = 1-p^{d/2} s, \,\,\,\,\, s = \frac{b}{a},\,\,\,\,\,\alpha=1+3\alpha_3,\,\,\,\,\,\beta=\alpha_3+4\alpha_4,\nonumber\\
	B \delta^i_j &=&X^i_j =-\left\{\(\frac{y+s-1}{s}\)(1+2\alpha y+3\beta y^2)+y(2+\alpha y)\right\}\delta^i_j.
\end{eqnarray}
For the scalar field, the equation of motion in Eq. \eqref{eom for phi} becomes
\begin{eqnarray}\label{eom for phi2}
	\nabla_\rho\nabla^\rho\phi-V_{,\phi}+\sqrt{\frac{d}{2(d+2)}}m_{pl}M^2_g\Phi=0.
\end{eqnarray}
In the same fashion, the constraint equation in Eq. \eqref{const-eq} can be expressed as
\begin{eqnarray}
	\(\nabla_\rho\nabla^\rho \phi-V_{,\phi}\) \nabla_\nu \phi &=& m^{2}_{pl}  \nabla_\rho \left(M^2_g X^\rho_{\nu}\right),
\end{eqnarray}
By using Eq. \eqref{eom for phi2} and then rewrite the equation in a convenient form, we have
\begin{eqnarray}\label{constraint}
	\sqrt{\frac{d}{2(d+2)}}\frac{\dot{\phi}}{m_{pl}H}=\frac{3(1-s)(A-B)}{(2A-\Phi)(1-s)+3s(A-B)},
\end{eqnarray}
where the over dot refers to the derivative with respect to $t$. From this constraint equation, an interesting branch of the solution is obtained $A=B$. For this branch, the equation of state parameter, $w_g$, is exactly equal to $-1$ and then the graviton mass will play the role of cosmological constant. Note that the energy density and pressure contributed from graviton mass can be respectively written as
\begin{eqnarray}
	\rho_g =  A m^2_{pl}M_g^2,\,\,\, p_g =-B m^2_{pl}M_g^2.
\end{eqnarray}
Considering the $A=B$ branch, called self-accelerating branch, one can see that the non-trivial solution is $\dot{\phi}/H = 0$. This means that the graviton mass, $M_g$, is constant. This behaviour cannot be obtained for MVMG since the graviton mass always decrease at late time.

Moreover, it is important to note that the equations of motion can be viewed as a coupling equations between a scalar field and the graviton mass. By rewriting equations of motion, the coupling equations can be written as
\begin{eqnarray}
	\rho'_\phi +3(1+w_\phi)\rho_\phi &=&\sqrt{\frac{d}{2(d+2)}}m_{pl}M^2_g\Phi\,\phi',\label{coup eq for phi}\\
	\rho'_g +3(1+w_g)\rho_g &=&-\sqrt{\frac{d}{2(d+2)}}m_{pl}M^2_g\Phi\,\phi',\label{coup eq for grav mass}
\end{eqnarray}
where
\begin{eqnarray}
	\rho_\phi = \frac{1}{2} \dot{\phi}^2+V,\,\,\,p_\phi = \frac{1}{2}\dot{\phi}^2-V,\,\,\, w_\phi =\frac{\frac{1}{2}\dot{\phi}^2-V}{\frac{1}{2}\dot{\phi}^2+V}.
\end{eqnarray}
and prime denotes the derivative with respect to $N = \ln a$. From these equations, it is found that the interaction term, which actually corresponds to the energy transfer between two contents, vanishes when $\phi' = 0$. This is consistent with the self-accelerating branch as we have discussed above. In the case of $\Phi=0$, the energy transfer between graviton mass and the scalar field also vanishes, 
but it is not necessary to obtain the solution which predicts the accelerating expansion of the universe because the solutions of $y$ which satisfy the case $A=B$ and $\Phi=0$ are different functions of the free parameters, $\alpha$ and $\beta$.
 Thus the condition $A=B$ is definitely enough to investigate the cosmological model in which there exists the fixed point which can explain the late-time expansion of the universe. Actually, for $\Phi = 0$, it does not correspond to the fixed point, then the scalar field still evolve. We will see these behavior more clearly when we consider the dynamical system and this is a main issue for the next section.


\section{dynamical system}\label{sec:dynamical system}
To see the behavior of each cosmological contents for the model of this effective massive gravity theory in more detail, we will analyze the model by using a dynamical system approach. As a result in the previous section, (0,0) components of Einstein equation in Eq. \eqref{Ein00} can be written as
\begin{eqnarray}\label{00 constraint}
	1
	&=&-\frac{A}{3}x+z^2+v+\Omega_m+\Omega_r,
\end{eqnarray}
where
\begin{eqnarray}
	x =-\frac{M_g^2}{H^2}, \,\,\,\, z^2 =\frac{\dot{\phi}^2}{6 m_{pl}^2 H^2},\,\,\,\,v=\frac{V}{3m^2_{pl}H^2},\,\,\,\, \Omega_m =\frac{\rho_m}{3 m_{pl}^2 H^2}, \,\,\,\, \Omega_r =\frac{\rho_r}{3 m_{pl}^2 H^2}.
\end{eqnarray}
This is one of the constraint equations which contains six variables including $y$. The others come from Eq. \eqref{constraint} which is expressed in terms of dynamical variables as
\begin{eqnarray}\label{z constraint}
	\sqrt{\frac{3d}{(d+2)}}z&=&\frac{3(A-B)(1-s)}{(2A-\Phi)(1-s)+3s(A-B)},\\
	A-B &=& \frac{1-s}{s} (y-1)Y,\,\,\,\,\,\,\,Y=3\beta y^2+2\alpha y +1.\label{eq-AmB}
\end{eqnarray}
It is important to note that the dynamics of the universe with the additional scalar field in this model is different from ones in the other usual scalar field models by virtue of the constraint equation,  Eq. \eqref{z constraint}, where $z \propto \phi'$. From this equation, the dynamics of scalar field is fixed by its own equation without coupling to the other dynamical variables as found in the usual scalar field models.

The third constraint equation is the consequence of the definition of the potential of the scalar field, $V(\phi)$ which is
\begin{eqnarray}\label{v constraint}
	v
	=\gamma x\(\frac{s}{1-y}\)^{\frac{4}{d}},\,\,\,\,\,\,\,\,\,\,\gamma=\frac{d(d-1)}{6}\frac{\kappa}{m^2_g}.
\end{eqnarray}
Note that we have introduced a new parameter $\gamma$ to characterize an effect of the potential term compared to the graviton mass. Actually, this parameter tells us how the curvature of the extra dimensions affects the dynamics of four-dimensional universe compared to the effect of the graviton mass. As a result, the geometry of the internal space can be characterized by this parameter as follows; the geometry of the  internal space is hyperbolic, flat and spherical for $\gamma < 0$, $\gamma = 0$ and $\gamma > 0$ respectively. Note also that for $d=1$, there is no internal curvature so that the potential term vanishes automatically.

Now we can choose to eliminate three dynamical variables, $z$, $v$ and $\Omega_m$ for convenience. As a result, three dynamical equations for three variables can be written as
\begin{eqnarray}
	x'&=&-x\(2\sqrt{\frac{3d}{d+2}}z+2\frac{\dot{H}}{H^2}\)\label{dyn of x},\\
	y'&=&\sqrt{\frac{3d}{d+2}}(y-1)z\label{dyn of y},\\
	\Omega'_r&=&-\Omega_r\(3(1+w_r)+2\frac{\dot{H}}{H^2}\)\label{dyn of Omega r},
\end{eqnarray}
where
\begin{eqnarray}
	w_{eff}=-1-\frac{2\dot{H}}{3H^2},\,\,\,\,\,\,\,\,\,\,
	 \frac{2\dot{H}}{3H^2}
	=\frac{x}{3}(A-B)-2z^2-(1+w_m)\Omega_m-(1+w_r)\Omega_r.
\end{eqnarray}
From this dynamical system, it is interesting to note that the equations do not depend on the number of the extra dimensions, $d$, after substituting $z$ from Eq. \eqref{z constraint}. Therefore, the dynamic of the contents in the universe is evolved by the same set of equations even though there are more number of the extra dimensions. However, the effect of the more higher dimensions than one is implicitly found in the constraint equation, Eq. \eqref{00 constraint}, where the potential term exists if $d > 1$.

By substituting $z$ from Eq. \eqref{z constraint} into Eq. \eqref{dyn of y}, one finds that the non-trivial fixed points occur at $A-B =0$. This equation can be solved for $y$ as
\begin{eqnarray}
	y_{\pm} = \frac{-\alpha \pm \sqrt{\alpha ^2-3 \beta } }{3 \beta }. \label{sol y+-}
\end{eqnarray}
As we have mentioned, these fixed points correspond to self-accelerating expansion of the universe. Note also that $x' =0$ in Eq. \eqref{dyn of x} is satisfied due to  $\dot{H}=0$. Moreover, from Eq. \eqref{dyn of y}, $y'$  depends only on $y$ so that one can examine the stability of the fixed points separately. As a result, the stability condition can be found by
\begin{eqnarray}
	(y-1)\partial_y z \Big|_{y=y_\pm} < 0.\label{stability}
\end{eqnarray}
From this condition, one can see that the quantity in the left hand side depends on three parameters; $\alpha$, $\beta$ and $s$. One of them can be eliminated by using the initial conditions such as $\Omega_{m0} \sim 0.25, \Omega_{r0} \sim 0 $ and $\Omega_{g0} = -A x/3+v \sim 0.75$. This can be done only in five-dimensional model where $v=0$. For the model with higher than five dimensions, we have one  more parameter, $\gamma$. Therefore, we will separate our analysis into two parts, five-dimensional model and  higher than five-dimensional model.

\subsection{Five-dimensional model}
\begin{figure}[h!]
\begin{center}
\includegraphics[scale=0.8]{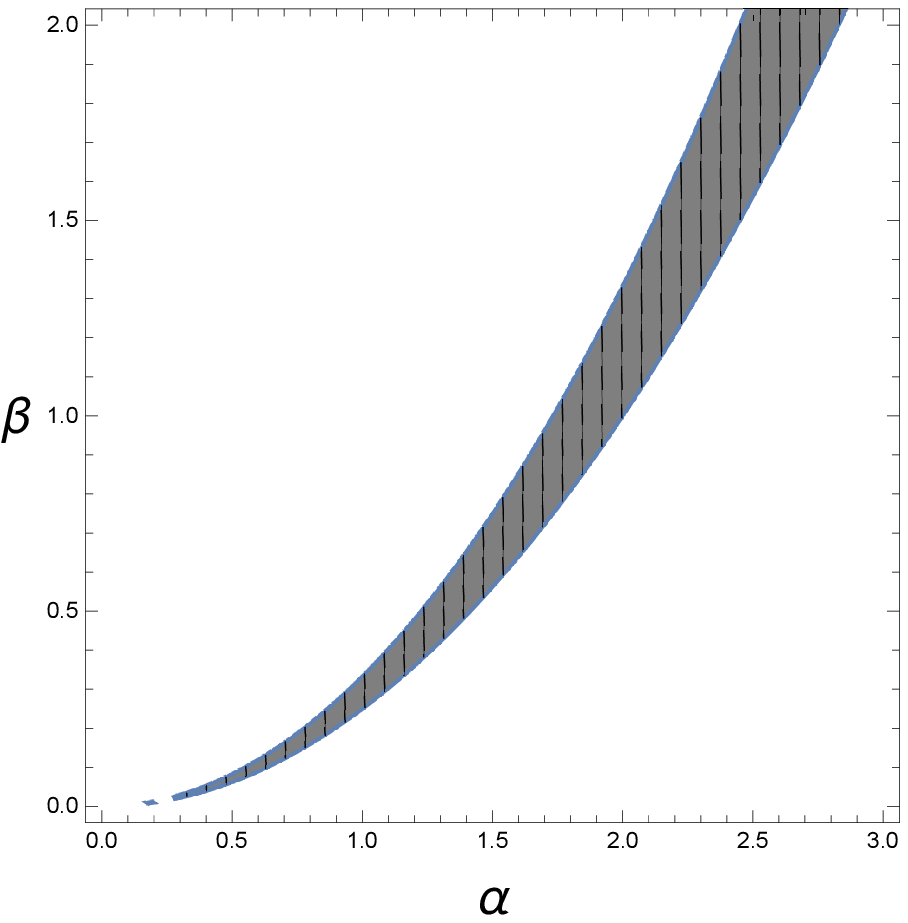}\qquad\qquad
\includegraphics[scale=0.8]{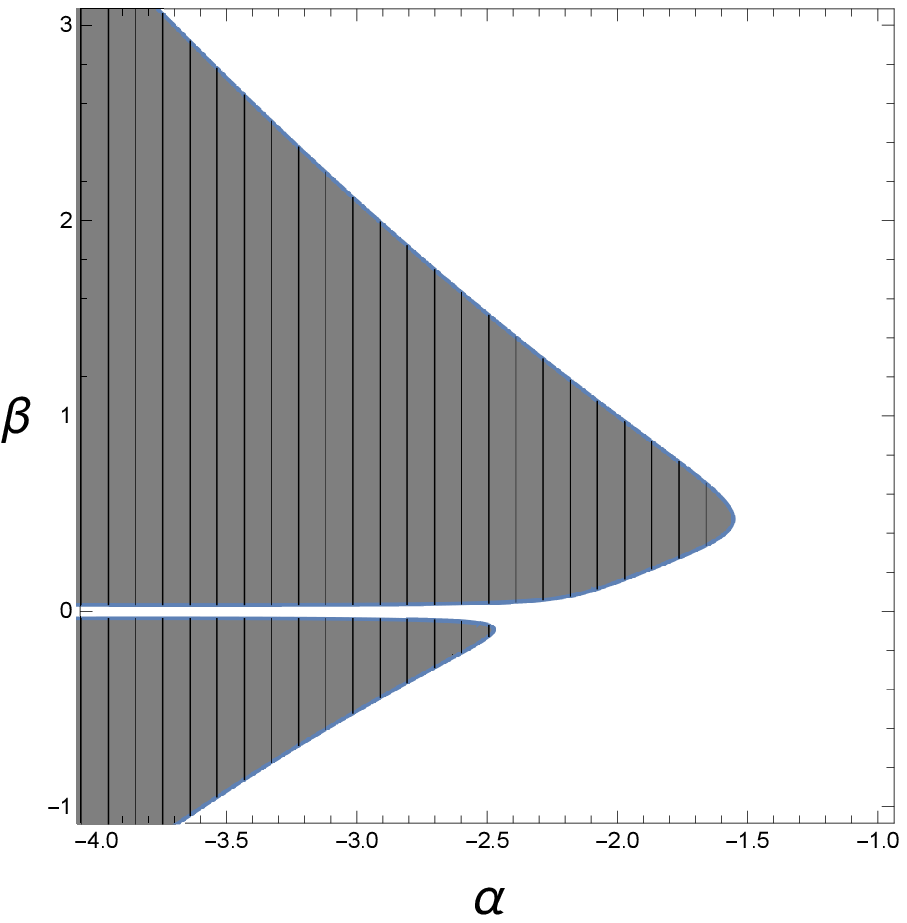}
\end{center}
{\caption{The left panel and the right panel show stability region for $y_{-}$ solution and $y_{+}$ solution respectively. }\label{fig:stability_0}}
\end{figure}

For five-dimensional model, the scalar potential vanishes then the constraint in Eq. (\ref{00 constraint}) becomes $-A x/3+\Omega_{m0} + \Omega_{r0} =1$. As we have mentioned above, one can find the stability condition in terms of parameter $(\alpha,\beta)$ by eliminating $s$ due to the initial conditions such that $\Omega_{m0} \sim 0.25, \Omega_{r0} \sim 0 $ and $\Omega_{g0} = -A x/3 \sim 0.75$. By using a numerical method, the regions satisfying the stability condition for $y_+$ and $y_-$ can be illustrated as in Fig. \ref{fig:stability_0}.

It is important to note that the stability condition in Eq. (\ref{stability}) can be rewritten explicitly by
\begin{eqnarray}
	\frac{1}{2A-\Phi}\partial_y Y\Big|_{y=y_\pm} < 0.\label{stability2}
\end{eqnarray}
Since solutions $y_{\pm}$ are solved from $Y=0$, $\partial_yY|_{y_+}$ is positive and $\partial_yY|_{y_-}$ is negative. Therefore, the stability condition can be inferred from the sign of $2A-\Phi$; $2A-\Phi<0$ for $y_+$ and $2A-\Phi>0$ for $y_-$. The regions illustrated in  Fig. \ref{fig:stability_0} include these conditions and conditions such that $y$ and $s$ are real number.

\begin{figure}[h!]
\begin{center}
\includegraphics[scale=0.6]{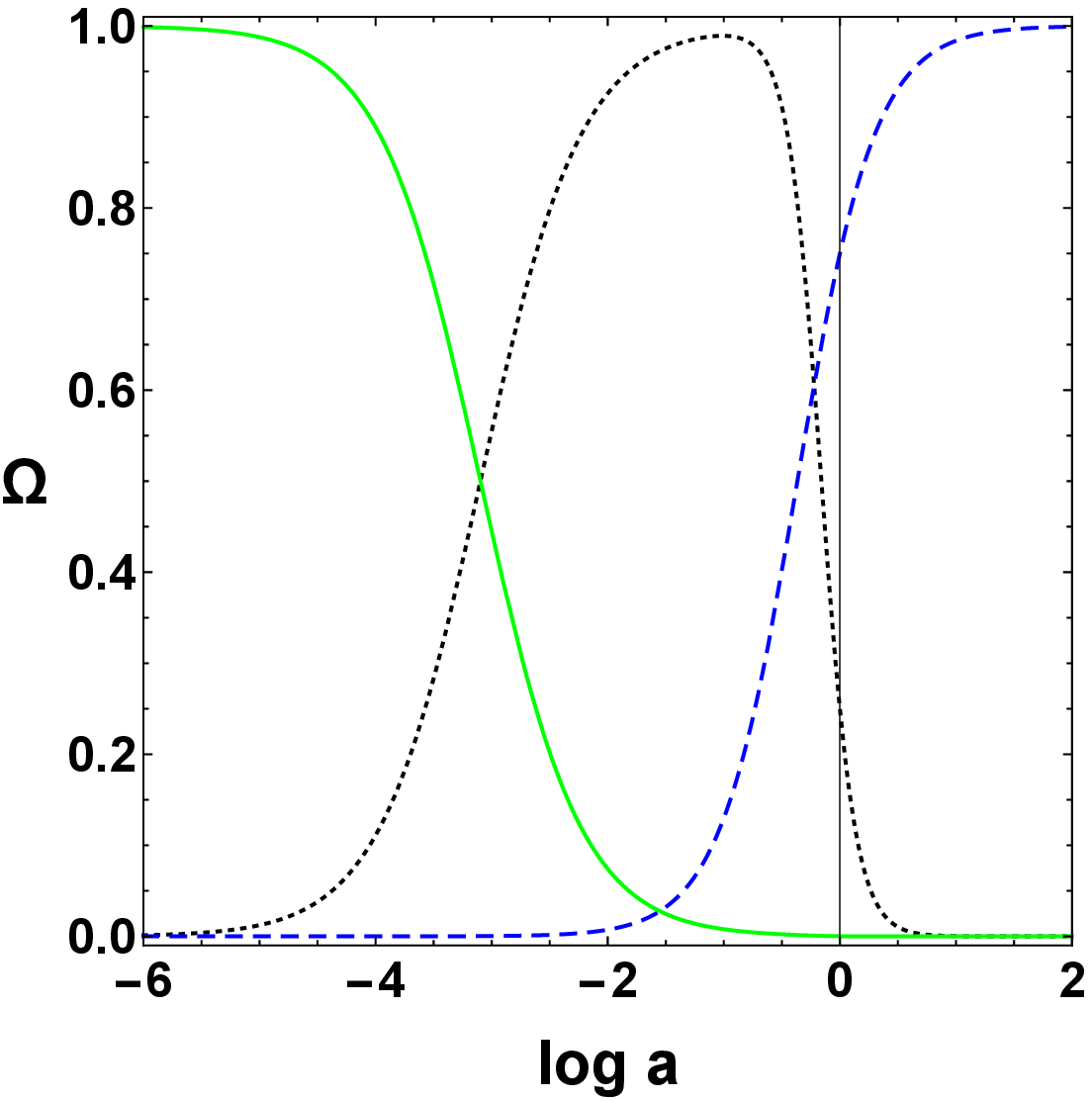}\qquad\qquad
\includegraphics[scale=0.6]{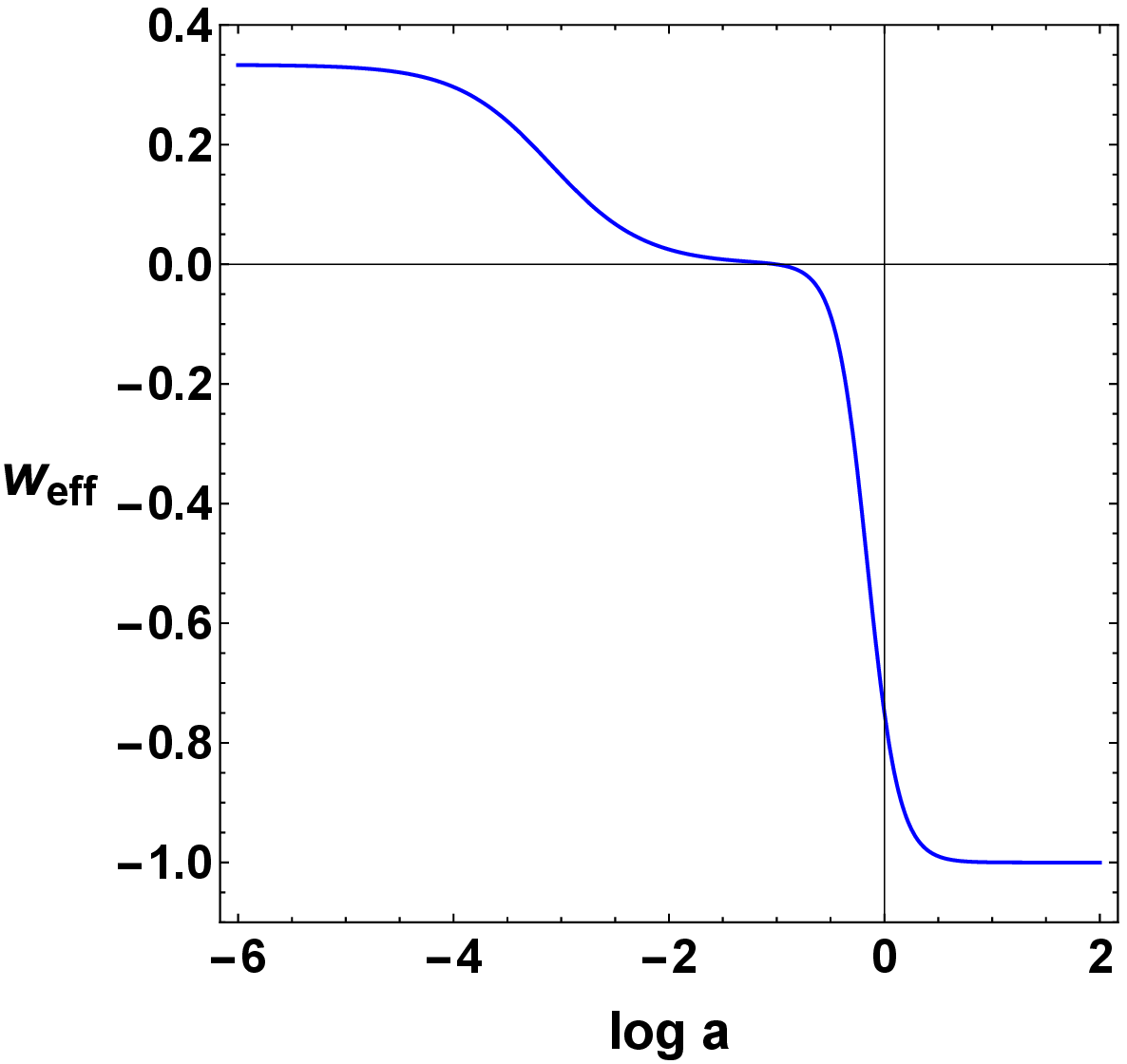}
\end{center}
{\caption{ The left panel shows the evolution of the density parameters of $\Omega_g$ (dashed-blue line), $\Omega_m$ (dotted-black line) and $\Omega_r$ (solid-green line).  The right panel shows the evolution of the $w_{eff}$. }\label{fig:evo5Ds}}
\end{figure}

There is a special case where $\partial_yY|_{y_{\mp}} = 0$. In this case, the eigenvalue is equal to zero and the stability cannot be inferred from linear analysis. Since this point corresponds to the minimum of $Y$, this guarantees the stability of the nonlinear analysis. As a result, the condition for this point can be written as $3\beta = \alpha^2$ which turns out that $y_+ = y_-$. Moreover, this point corresponds to the case $z=0$ and $z'=0$. This means that the scalar field is fixed all the time of the evolution of the universe. The radius of the extra dimension is also stabilized at this fixed point. By choosing $\alpha = 2$, the evolution of the contents in the universe, $\Omega_g, \Omega_m, \Omega_r$  and the $w_{eff}$ can be numerically evaluated as shown in Fig. \ref{fig:evo5Ds}. From this figure, one can see that all contents evolve as in standard evolution such that there exist the radiation, matter and dark energy dominated periods.

Now we proceed our analysis to the case of  $\partial_yY|_{y_{\mp}} \neq 0$. For this case, it is expected that the dynamics of the scalar field will affect the evolution of the universe since $\partial_y z \neq 0$ leading to $z'\neq0$.

\begin{figure}[h!]
\begin{center}
\includegraphics[scale=0.6]{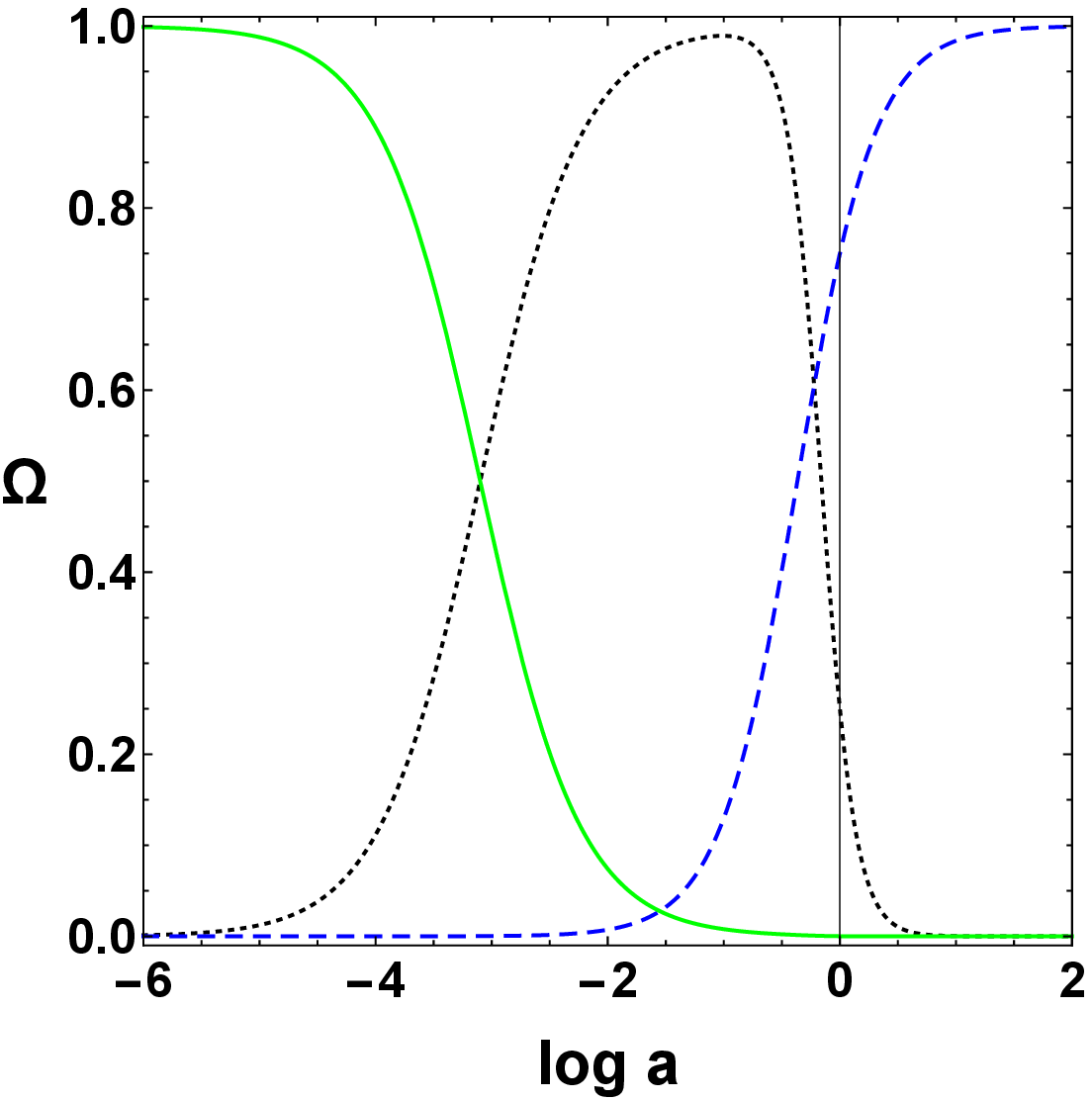}\qquad\qquad
\includegraphics[scale=0.6]{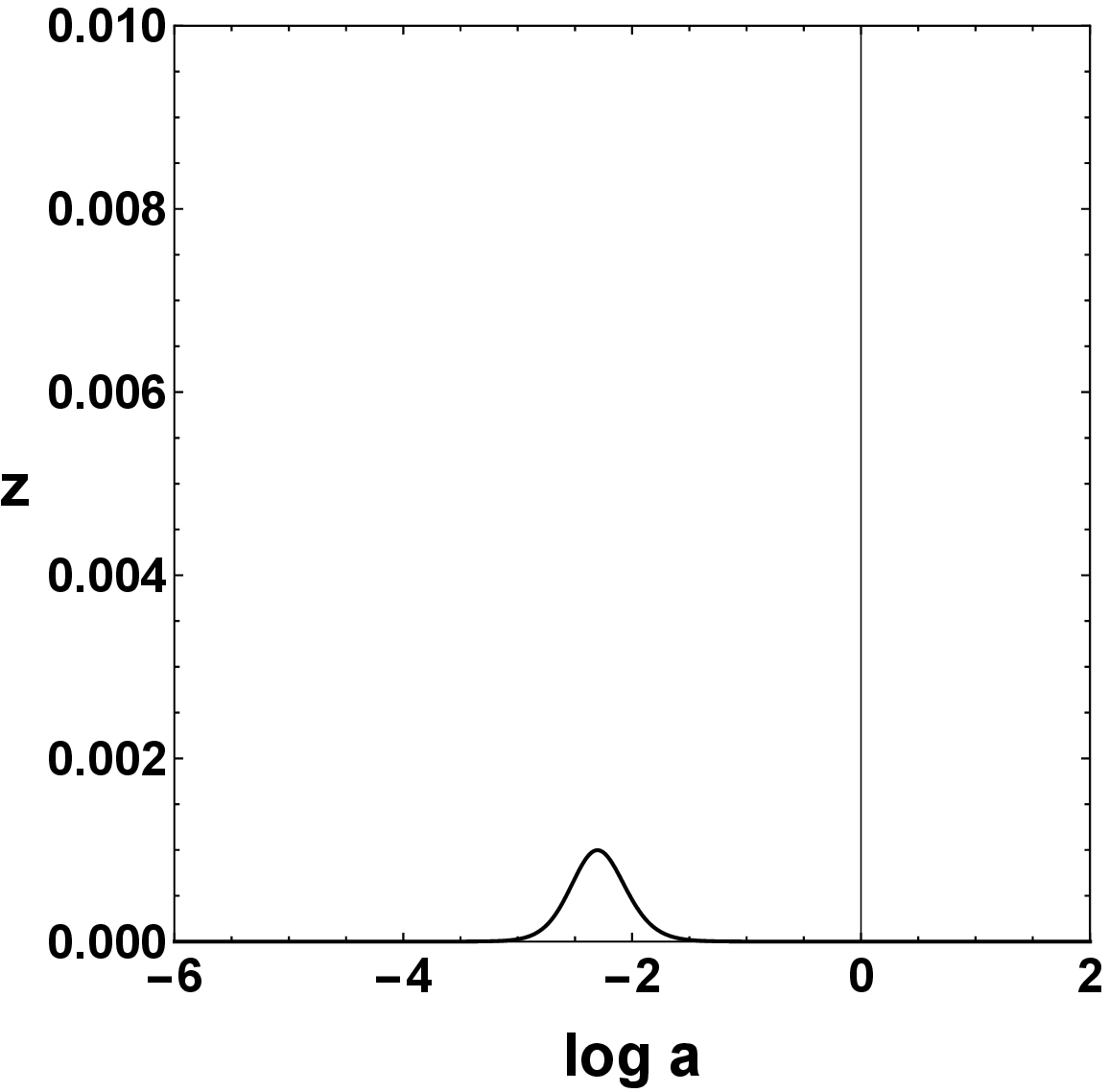}
\end{center}
{\caption{ The left panel shows the evolution of the density parameters of $\Omega_g$ (dashed-blue line), $\Omega_m$ (dotted-black line) and $\Omega_r$ (solid-green line).  The right panel shows the evolution of the $z$. }\label{fig:evo5Dym}}
\end{figure}


For $y_-$ solution, the stability region in Fig \ref{fig:stability_0} is a region below the line $\alpha^2=3\beta, \alpha > 0$. Since $w_\phi =1$, so that if $z$ takes the value of order $0.1$, $w_{eff}$ will significantly deviate from the standard evolution. In order to obtain such a small effect of the kinetic of the scalar field, one has to take the parameter slightly from $\alpha^2=3\beta$ line. As a result, we choose $\alpha = 2, \beta = 1.32$ and then the evolution of the contents in the universe is shown in Fig. \ref{fig:evo5Dym}. From this figure, one found that there is a peak of $z$ about $z\sim 0.001$ at $\log a \sim -2$. This peak takes a small value due to choosing parameters $\alpha,\beta $ as discussed above. Note that the peak can be eliminated by choosing small value of $\alpha, \beta$, for example $\alpha = 0.5, \beta=0.0825$, leading to small value of $\partial z /\partial y$. Though it is not shown here, we demonstrate that we can obtain the standard evolution by setting as discussed.

For $y_+$ solution, the stability region is larger than $y_-$ solution. A behavior of the density parameters for each content in the universe is similar to ones in $y_-$ solution. In order to obtain the standard evolution of the density parameters $-$ $z$ does not significantly emerge during the whole predicted evolution $-$ one has to choose the parameter $\alpha, \beta$, for example $\alpha = -2.0, \beta=0.5$,  such that $\partial z /\partial y \geqslant -0.1$. Since the evolution is similar to $y_-$ solution, we do not show it explicitly in this presentation.

It is important to note that $z$ may dominate at early times since $\partial z /\partial y  \neq 0$. In this case, one has to tune parameters such that  $\partial z /\partial y \geqslant -0.01$. For this condition, $z$ still take the value less than $10^{-4}$ at $\log a \sim -50$. This is enough to guarantee that the change of the radius of the extra dimension does not affect the Big Bang Nucleosynthesis.


\subsection{Six-dimensional model}

In this subsection, we will consider the dynamics of the universe including the potential term of the scalar field. As we have mentioned, the effect of the potential term can be characterized by parameter $\gamma$ corresponding to the curvature of the internal space. One of important issues in this model is that the potential term is not involved in the dynamical system explicitly and then its dynamics is completely determined since it can be written in terms of $y$. Therefore, the effect of this term is concerned mostly in the region of stability of the fixed point.

\begin{figure}[h!]
\begin{center}
\includegraphics[scale=0.8]{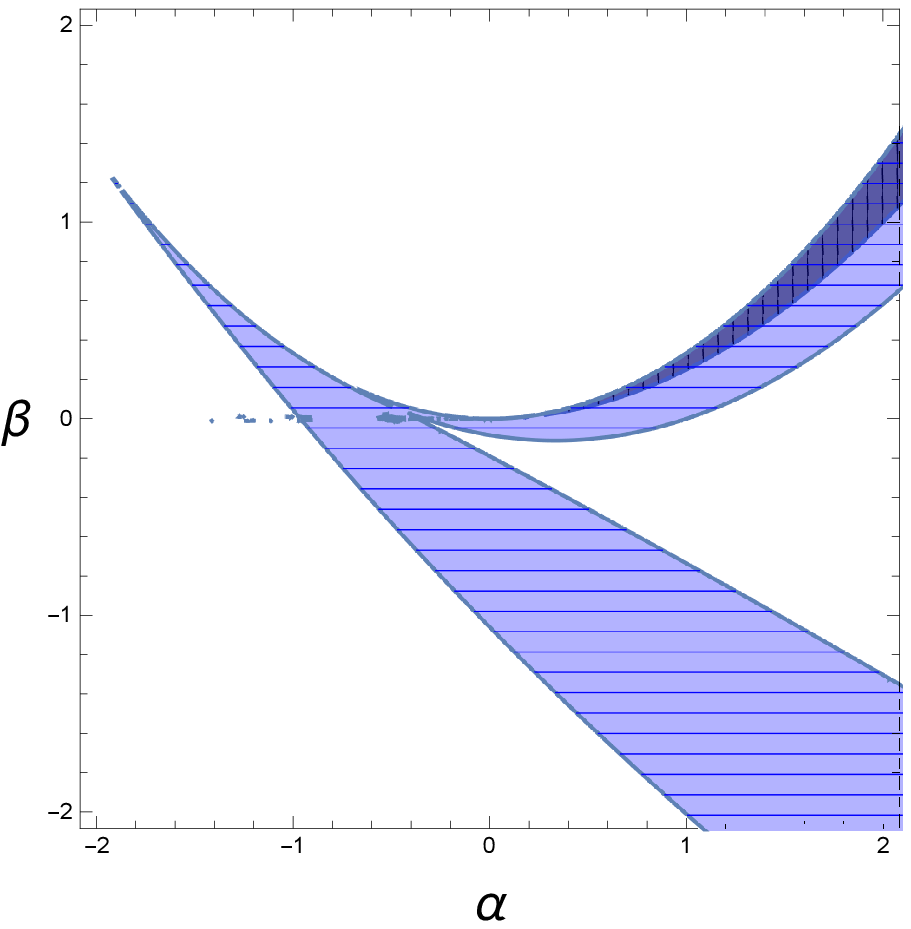}\qquad\qquad
\includegraphics[scale=0.8]{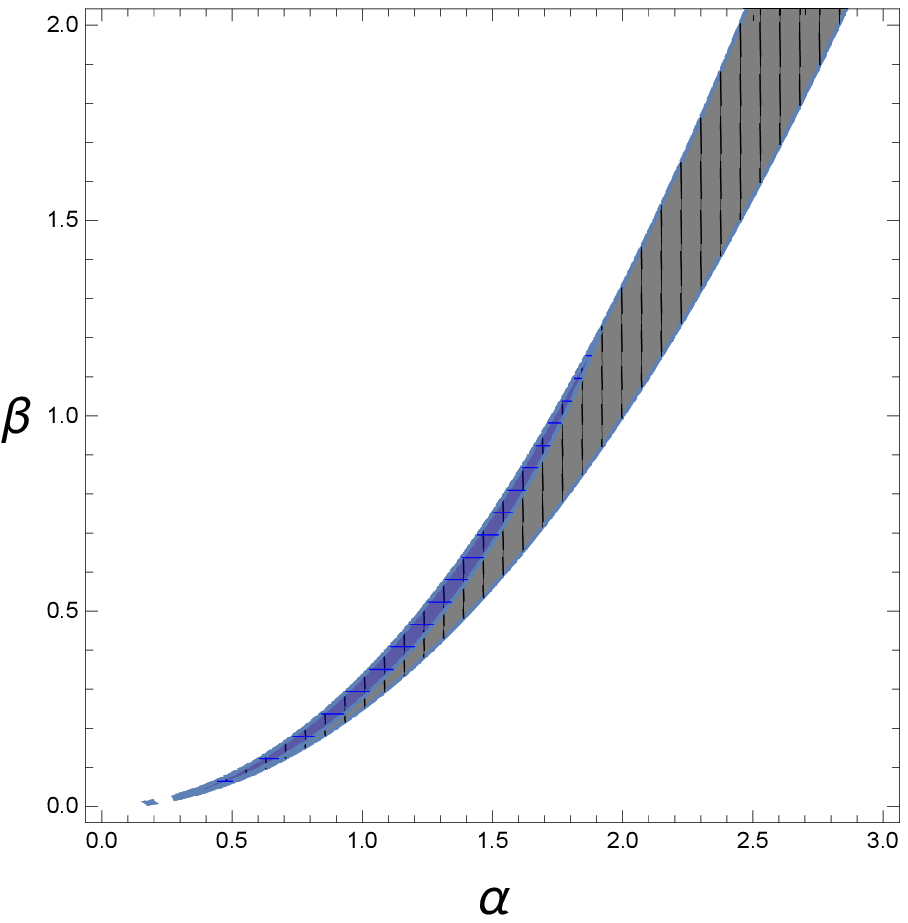}
\end{center}
{\caption{The left panel and the right panel show stability region for $y_{-}$ solution with $\gamma=-0.1$ and $0.01$ respectively. The horizontal-blue-shaded region corresponds to the stability region in 6D model while the vertical-black shaded region corresponds to the stability region in 5D model. }\label{fig:stability_ym}}
\end{figure}

As we have discussed in five dimensions case, there exists a special class of the solution such that $y_-=y_+$ (or in terms of parameters $\beta = \alpha^2/3$). Therefore, it is instructive to consider this class of the solution to find out how the parameter affect the dynamics. As a result, the parameter $s$ which determined from the initial value can be expressed as
\begin{eqnarray}
s^2 = \frac{6(1+\alpha)^2(1-\Omega_{m0})}{\alpha(1\pm\sqrt{1-36\alpha^2 \gamma (1-\Omega_{m0})})}.\label{s2}
\end{eqnarray}
In order to obtain the real value of $s$, the parameter $\gamma$ is restricted by
\begin{eqnarray}
\gamma \leq \frac{1}{36\alpha^2  (1-\Omega_{m0})}.\label{gamma-con}
\end{eqnarray}
From this condition, one finds that $\gamma$ cannot take very large positive value without tuning parameter $\alpha$. Actually, for order unity of $\alpha$, $\gamma$ can take order of $10^{-2}$. This will make the stability region narrow for positive value $\gamma$. On the other hand, $\gamma$ can take very large negative value. This provide larger region of stability than one for $\gamma=0$. The stability regions for $y_-$ solution with $\gamma = 0.01$ and $\gamma=-0.1$ are shown in Fig. \ref{fig:stability_ym}. From this figure, one can see that for negative $\gamma$, the stability region is enlarged while the region is reduced for positive $\gamma$. Note that for $y_+$ solution, the change of the region due to the existence of $\gamma$ is similar to one for $y_-$. We do not show explicitly in this presentation.

\begin{figure}[h!]
\begin{center}
\includegraphics[scale=0.8]{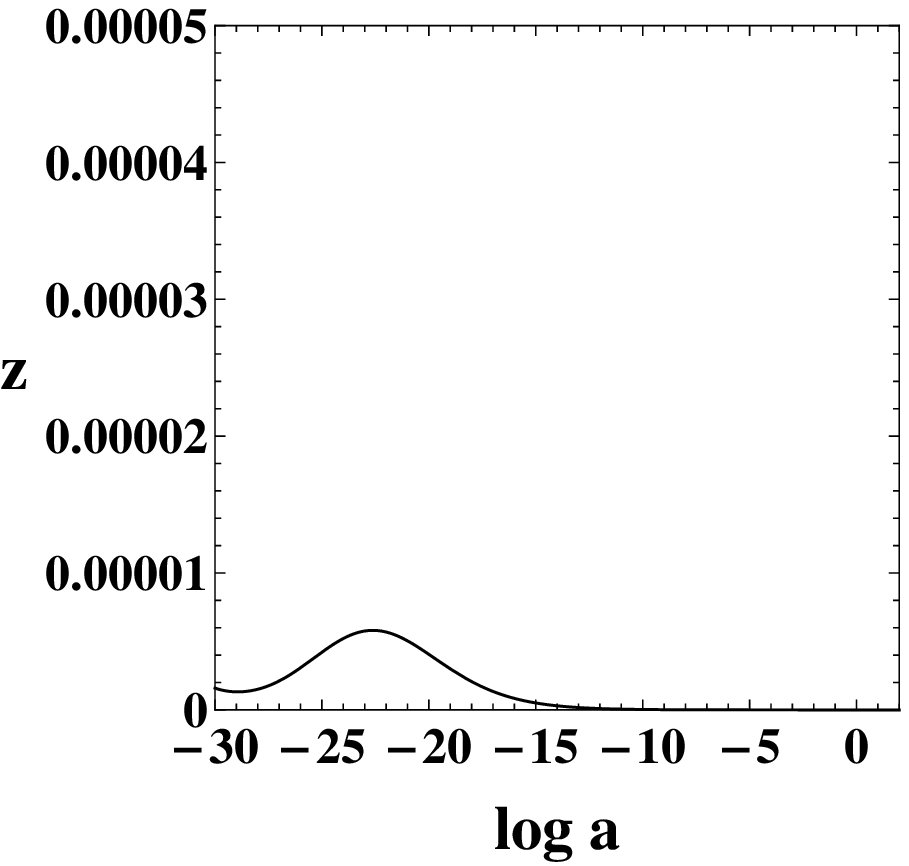}\qquad\qquad
\includegraphics[scale=0.62]{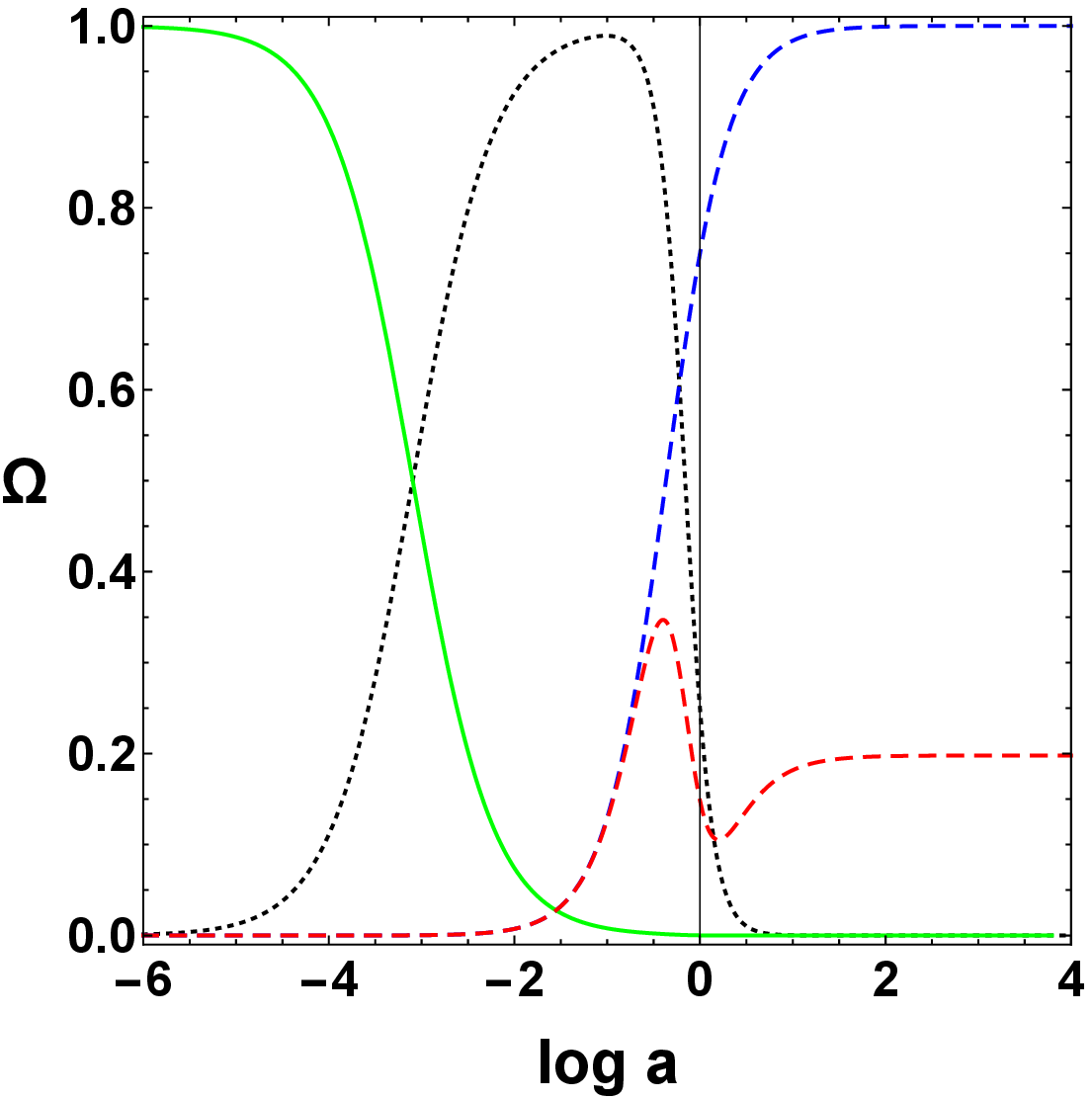}
\end{center}
{\caption{The left panel shows evolution of $z$ for $y_{-}$ solution with $\gamma=-2.0, \alpha = 2, \beta = 1.32$. The right panel shows evolution of the density parameters of $\Omega_g$ (dashed-blue line), $v$ (dashed-red line),$\Omega_m$ (dotted-black line) and $\Omega_r$ (solid-green line) for $y_{+}$ solution with $\gamma=-2.0, \alpha = -2, \beta = 0.5$. }\label{fig:evo6D}}
\end{figure}

For the evolution of the density parameters, the behavior can be seen by using Eq. (\ref{s2}). From this equation, it is found that the existence of $\gamma$ can make $s$ decreased and $\partial_y z \propto (s-1)^2$. Therefore the dynamics of $z$ is reduced and then the peak reduces and moves to the earlier time. This behavior is confirmed by using numerical calculation as shown in the left panel of Fig. \ref{fig:evo6D}. By comparing this figure to the right panel of Fig. \ref{fig:evo5Dym}, one can see that the existence of potential term will reduce the peak of $z$ and shift the peak to earlier time. We also found this behavior in both $y_+$ and $y_-$ solutions. However, it is not obvious in the case of $y_+$ since the stability region takes place in $\alpha <0$ region.

One more feature of the dynamics in six-dimensions is that the contribution to drive the late-time expansion of the universe does not only come from the graviton mass but also from the potential term of scalar field. In the right panel of Fig. \ref{fig:evo6D}, the evolution of the contents in the universe is illustrated including the contribution from scalar potential, $v$ (dashed-red line). In this case, we found that the expansion of the universe nowadays is driven by the graviton mass and the curvature of the extra dimension.   

Note that for the case $s=1$, $y$ will be a non-dynamical variable then the scalar field is just a constant in the dynamical system. Therefore, this case gives us the same prediction as GR with the cosmological constant.

It is worthwhile to note that one of the problem from the cosmological model reduced from the higher dimensional space time is that it is difficult to find the mechanism to stabilize the extra dimensions. For example, in the Casimir dark energy models, one have to add other structure, Aether field \cite{Chatrabhuti:2009ew} or Gauss-Bonnet terms \cite{Wongjun:2013jna}, to stabilize the extra dimensions. In our case, without the graviton mass term, the scalar potential is the exponentially decay so that the scalar field roll the potential down to the very large value. Ultimately, the extra dimensions cannot be compactified. However, by including the graviton mass terms, we found that the extra dimensions can be stabilized so that it is possible to compactified the extra dimensions to small value for hiding it from the observations. This is one of important result of the model.

\section{conclusions} \label{sec:conclusion}

As we have mention, the dRGT massive gravity theory with the Minkowski fiducial metric admit only the open FLRW solution. To obtain all kinds of FLRW solutions, one may use the other form of this metric which is flat FLRW form. Unfortunately, the further problem about the lack of the number of degrees of freedom is arisen. To solve this problem the external degrees of freedom is introduced to this massive gravity theory. In this work, we perform the Kaluza-Klein dimensional reduction for the $(4+d)$-dimensional dRGT massive gravity theory. By our ansatz, the extra dimensions are assumed to be maximally symmetric. The off-diagonal terms in the ansatz metric do not appear in order to provide only a scalar field in the four dimensional effective massive gravity theory. The  radius of the extra dimensions are also interpreted as the function of the scalar field.  We found that this effective massive gravity theory in four-dimensional spacetime contain description of two extensions of dRGT theory, namely mass-varying massive gravity (MVMG) and quasidilaton massive gravity (QMG). Up to our knowledge, this kind of extended dRGT massive gravity has not been investigated yet. Therefore, we then investigate the cosmological model due to this  effective massive gravity theory.

In order to investigate the cosmological solutions from our effective massive gravity theory, we adopt the form of physical metric and fiducial metric as the FLRW metric form. In order to complete our analysis, matter and radiation are introduced into our cosmological model. It is found that the graviton mass and the scalar field are coupled. We also found that it is possible to obtain solutions providing the self-accelerating expansion of the universe (the case $A=B$). We pay attention to this possibility in detail by using the dynamical system approach.

To analyze dynamics of these cosmological models, the six dynamical variables $x, y, z, v, \Omega_m$ and $\Omega_r$ with four free parameters, $\alpha, \beta, s$  and $\gamma$ are defined. For the five-dimensional model, in the special case where $y_-=y_+$, the kinetic term of the scalar field, or variable $z$, vanishes exactly. The free parameters $\alpha$ and $\beta$ can be chosen without regarding the effect of $z$. However, the effect of $z$ must be small for the case of $y_-\neq y_+$ in order to obtain the expected evolution of the universe.

For the higher-dimensional model which is studied as six dimensions in this work, the potential of the scalar field characterized by the free parameter $\gamma$ is introduced. We found that dynamics of $z$ is affected by this free parameter through the ratio of two scale factors, $s$ and the existence of the scalar potential provides us a larger stability region of $\alpha$ and $\beta$. The appropriated free parameters which give the standard evolution of the universe for each part of consideration have already been discussed (see in our Section \ref{sec:dynamical system}). The crucial difference between the five-dimensional model and the higher one is the late-time expansion of the universe is obtained due to only the graviton mass in five dimensional spacetime while the both of graviton mass and the potential of the scalar field can drive the accelerating expansion of the universe. Moreover, by including the graviton  mass term, we found that the extra dimensions can be stabilized. Therefore, it is possible to compactified the extra dimensions to small value for hiding it from the observations while it is not possible for the case such that the graviton mass is not introduced. This may shed light on the interplay between the consistent cosmological model and the fundamental theory with higher-dimensional spacetime.

Note that this effective massive gravity theory is not studied in the general case which is the case of $r\neq 0$ or $p\neq q$. Therefore, the cosmological implication in more complicated case is still the interesting challenge. The instability issue as well as the number of degree of freedom of the theory is one of interesting issues to investigate. We leave this investigation for further work.  

In our investigation, the matter is not coupled to the scalar field. However, in principle, it is possible that they are coupled together since the conformal factor will influence matter Lagrangian. This may lead to coupling model between dark energy contributed from graviton mass and the matter field. Therefore it may be possible to use this coupling model to examine the possible way to solve the coincidence problem. We leave this investigation for further work.


\begin{acknowledgements}
We would like to thank Lunchakorn Tannukij for value discussion and reading through the manuscript. This project is partially supported by the ICTP through grant No. OEA-NT-01. 
\end{acknowledgements}


\begin{appendix}

\section{The explicit form of $\tilde{U}$}\label{app:Ut}
To construct the massive gravity theory which eliminates BD ghost, each potential $\tilde{U}_i$ is implied from the characteristic equation given $i\times i$ square matrix $\mathcal{\tilde{K}}$, $\mathcal{\tilde{K}}_{i\times i}$ \cite{Do:2016abo}. In this work, we use $\tilde{K}$ denotes $\mathcal{\tilde{K}}_{i\times i}$ in which their determinants can be constructed $\tilde{U}_i$ by $\text{det}~\mathcal{\tilde{K}}_{i\times i}=\text{det}~\tilde{K}=\frac{1}{2 (i)!}\tilde{U}_i$. The form of $\tilde{U}_i$ in Eq. \eqref{4+d action} will be expressed up to $i=6$ as
\begin{eqnarray}
	{\tilde U}_2&=&[{\tilde K}]^2-[{\tilde K}^2], \nonumber\\
	{\tilde U}_3&=&[{\tilde K}]^3-3[{\tilde K}][{\tilde K}^2]+2[{\tilde K}^3], \nonumber\\
	{\tilde U}_4&=&[{\tilde K}]^4-6[{\tilde K}]^2[{\tilde K}^2]+3[{\tilde K}^2]^2+8[{\tilde K}][{\tilde K}^3]-6[{\tilde K}^4],  \nonumber\\
	{\tilde U}_5&=&[{\tilde K}]^5-10[{\tilde K}]^3[{\tilde K}^2]+20[{\tilde K}]^2[{\tilde K}^3]-20[{\tilde K}^2][{\tilde K}^3]+15[{\tilde K}][{\tilde K}^2]^2-30[{\tilde K}][{\tilde K}^4]+24[{\tilde K}^5], \nonumber\\
	{\tilde U}_6&=&[{\tilde K}]^6-15[{\tilde K}]^4[{\tilde K}^2]+40[{\tilde K}]^3[{\tilde K}^3]-90[{\tilde K}]^2[{\tilde K}^4]+45[{\tilde K}]^2[{\tilde K}^2]^2
	-15[{\tilde K}^2]^3\nonumber\\
			 &&+40[{\tilde K}^3]^2-120[{\tilde K}^3][{\tilde K}^2][{\tilde K}]+90[{\tilde K}^4][{\tilde K}^2]+144[{\tilde K}^5][{\tilde K}]-120[{\tilde K}^6].
\end{eqnarray}

\section{The explicit form of $F$}\label{app:F}
The additional potential obtained from the dimensional reduction was defined in Eq. \eqref{extend 4+d action} where
\begin{eqnarray}
	F=F_2+\alpha_3F_3+\alpha_4F_4+\alpha_5F_5+\alpha_6F_6,
\end{eqnarray}
with
\begin{eqnarray}
	F_2&=&2[K]+r(d-1),\nonumber\\
	F_3&=&3U_2+3r(d-1)[K]+r^2(d-1)(d-2),\nonumber\\
	F_4&=&4U_3 +6r(d-1)U_2+4r^2(d-1)(d-2)[K]+r^3(d-1)(d-2)(d-3),\nonumber\\
	F_5&=&5U_4+10r(d-1)U_3+6r^2(d-1)(d-2)U_2+5r^3(d-1)(d-2)(d-3)[K]\nonumber\\
		&&+r^4(d-1)(d-2)(d-3)(d-4),\nonumber\\
	F_6&=&6U_5+15r(d-1)U_4+20r^2(d-1)(d-2)U_3+15r^3(d-1)(d-2)(d-3)U_2\nonumber\\
		&&+6r^4(d-1)(d-2)(d-3)(d-4)[K]+r^5(d-1)(d-2)(d-3)(d-4)(d-5).\nonumber\\
\end{eqnarray}

\section{The explicit forms of $X_{\mu\nu}$ and $Y_{\mu\nu}$}\label{app:Xuv Yuv}
By varying the mass terms in the action in Eq. \eqref{4 action} with respect to the metric $g_{\mu\nu}$ which were shown in Eq. \eqref{eom for guv}, we obtained two tensors $X_{\mu\nu}$ and $Y_{\mu\nu}$ explicitly written as
\begin{eqnarray}
	 X_{\mu\nu}=X^{(2)}_{\mu\nu}+\alpha_3X^{(3)}_{\mu\nu}+\alpha_4X^{(4)}_{\mu\nu}+\alpha_5X^{(5)}_{\mu\nu}+\alpha_6X^{(6)}_{\mu\nu},
\end{eqnarray}
with
\begin{eqnarray}
	X^{(2)}_{\mu\nu}&=&-K^2_{\mu\nu}+([K]+1) K_{\mu\nu}-\frac{1}{2}(U_2+2[K])g_{\mu\nu},\nonumber\\
	X^{(3)}_{\mu\nu}&=&\frac{1}{2} \left\{6K^3_{\mu\nu}-6([K]+1)K^2_{\mu\nu}+3(U_2+2[K])K_{\mu\nu}-(U_3+3U_2)g_{\mu\nu}\right\},\nonumber\\
	X^{(4)}_{\mu\nu}&=&2\left\{-6K^4_{\mu\nu}+6([K]+1)K^3_{\mu\nu}-3(U_2+2[K])K^2_{\mu\nu}+(U_3+3U_2)K_{\mu\nu}\right\}-\frac{1}{2}(U_4+4U_3)g_{\mu\nu},\nonumber\\
	X^{(5)}_{\mu\nu}&=&\frac{1}{2}\left[5\left\{\begin{array}{l}
	24K^5_{\mu\nu}-24([K]+1)K^4_{\mu\nu}+12(U_2+2[K])K^3_{\mu\nu}
	\\-4(U_3+3U_2)K^2_{\mu\nu}+(U_4+4U_3)K_{\mu\nu}\end{array}\right\}
	-(U_5+5U_4)g_{\mu\nu}\right],\nonumber\\
	X^{(6)}_{\mu\nu}&=&3\left\{\begin{array}{l}
	-120K^6_{\mu\nu}+120([K]+1)K^5_{\mu\nu}-60(U_2+2[K])K^4_{\mu\nu}\\
	+20(U_3+3U_2)K^3_{\mu\nu}-5(U_4+4U_3)K^2_{\mu\nu}+(U_5+5U_4)K_{\mu\nu}\end{array}\right\}
	-\frac{1}{2}(U_6+6U_5)g_{\mu\nu},\nonumber\\
\end{eqnarray}
and 
\begin{eqnarray}
	Y_{\mu\nu}=Y^{(2)}_{\mu\nu}+\alpha_3Y^{(3)}_{\mu\nu}+\alpha_4Y^{(4)}_{\mu\nu}+\alpha_5Y^{(5)}_{\mu\nu}+\alpha_6Y^{(6)}_{\mu\nu},
\end{eqnarray}
with
\begin{eqnarray}
	Y^{(2)}_{\mu\nu}&=&
	\frac{1}{2}\[2\left\{K_{\mu\nu}-([K]+1)g_{\mu\nu}\right\}
	+r(d-1)g_{\mu\nu}\],\nonumber\\
	Y^{(3)}_{\mu\nu}&=&
	\frac{1}{2} \[\begin{array}{l}-3\left\{2 K^2_{\mu\nu}-2 ([K]+1) K_{\mu\nu}+(U_2+2 [K])g_{\mu\nu} \right\}
	+3 r (d-1)  \left\{K_{\mu\nu}-g_{\mu\nu} ([K]+1)\right\}\\
	-r^2(d-1) (d-2) g_{\mu\nu}\end{array}\],\nonumber\\
	Y^{(4)}_{\mu\nu}&=&
	\frac{1}{2}\[\begin{array}{l}4\left\{6 K^3_{\mu\nu}-6 ([K]+1) K^2_{\mu\nu}+3(U_2+2 [K])K_{\mu\nu}-(U_3+3 U_2)g_{\mu\nu} \right\}\\
	-6 r (d-1) \left\{2 K^2_{\mu\nu}-2 ([K]+1) K_{\mu\nu}+(U_2+2[K])g_{\mu\nu} \right\}\\
	+4 r^2(d-1) (d-2)   \left\{K_{\mu\nu}-([K]+1)g_{\mu\nu} \right\}
	-r^3(d-1) (d-2) (d-3) g_{\mu\nu} \end{array}\],\nonumber\\
	Y^{(5)}_{\mu\nu}&=&
	\frac{1}{2} \[\begin{array}{l}-5\left\{24K^4_{\mu\nu}-24([K]+1) K^3_{\mu\nu}+12 (U_2+2[K])K^2_{\mu\nu}-4(U_3+3 U_2)K_{\mu\nu}+(U_4+4 U_3)g_{\mu\nu} \right\}\\
	+10r (d-1) \left\{6 K^3_{\mu\nu}-6 ([K]+1) K^2_{\mu\nu}+3 (U_2+2 [K])K_{\mu\nu}-(U_3+3 U_2)g_{\mu\nu} \right\}\\
	-10 r^2(d-1) (d-2) \left\{2 K^2_{\mu\nu}-2 ([K]+1) K_{\mu\nu}+(U_2+2 [K])g_{\mu\nu} \right\}\\
	+5 r^3(d-1) (d-2) (d-3)  \left\{K_{\mu\nu}-([K]+1)g_{\mu\nu} \right\}
	-r^4(d-1) (d-2) (d-3) (d-4)g_{\mu\nu}\end{array}\],\nonumber
\end{eqnarray}
\begin{eqnarray}
	Y^{(6)}_{\mu\nu}&=&
	 \frac{1}{2}\[\begin{array}{l}6\left\{\begin{array}{l}120K^5_{\mu\nu}-120([K]+1)K^4_{\mu\nu}+60(U_2+2[K])K^3_{\mu\nu}-20(U_3+3U_2)K^2_{\mu\nu}\\+5(U_4+4U_3)K_{\mu\nu}-(U_5+5U_4)g_{\mu\nu} \end{array}\right\}\\
	-15 r(d-1) \left\{\begin{array}{l}24K^4_{\mu\nu}-24([K]+1)K^3_{\mu\nu}+12(U_2+2[K])K^2_{\mu\nu}-4(U_3+3U_2)K_{\mu\nu}\\
	+(U_4+4U_3)g_{\mu\nu} \end{array}\right\}\\
	+20 r^2(d-1) (d-2) \left\{6K^3_{\mu\nu}-6 ([K]+1)K^2_{\mu\nu}+3(U_2+2[K])K_{\mu\nu}-(U_3+3U_2)g_{\mu\nu} \right\}\\
	-15 r^3(d-1) (d-2) (d-3) \left\{2K^2_{\mu\nu}-2 ([K]+1) K_{\mu\nu}+(U_2+2[K])g_{\mu\nu} \right\}\\
	+6 r^4(d-1) (d-2) (d-3) (d-4) \left\{K_{\mu\nu}-([K]+1)g_{\mu\nu} \right\}\\
	-r^5(d-1) (d-2) (d-3) (d-4)(d-5) g_{\mu\nu} \end{array}\].\nonumber\\
\end{eqnarray}

\section{The explicit forms of $\Phi$ and $\Psi$}\label{app:Phi Psi}

In the equation of motion for $\phi$ which was shown in Eq. \eqref{eom for phi}, the scalar functions, $\Phi$ and $\Psi$ can be written as
\begin{eqnarray}
	\Phi=m_{pl}\sqrt{\frac{d+2}{2d}}\frac{\delta U}{\delta\phi}-U=\Phi_2+\alpha_3\Phi_3+\alpha_4\Phi_4+\alpha_5\Phi_5+\alpha_6\Phi_6,
\end{eqnarray}	
with
\begin{eqnarray}
	\Phi_2=-3[K],\,\,\,\Phi_3=\frac{1}{2}U_3-3U_2,\,\,\,\Phi_4=U_4-2U_3,\,\,\,\Phi_5=\frac{3}{2}U_5,\,\,\,\Phi_6=2U_6+3U_5,
\end{eqnarray}
and 
\begin{eqnarray}
	\Psi=m_{pl}\sqrt{\frac{d+2}{2d}}\frac{\delta F}{\delta\phi}-F=\Psi_2+\alpha_3\Psi_3+\alpha_4\Psi_4+\alpha_5\Psi_5+\alpha_6\Psi_6,
\end{eqnarray}	
with
\begin{eqnarray}
	\Psi_2&=&-([K]+4)-r(d-1),\nonumber\\
	\Psi_3&=&-9[K]-\frac{3}{2} r(d-1)([K]+4)-r^2 (d-1)(d-2),\nonumber\\
	\Psi_4&=&2(U_3-6U_2)-18r(d-1)[K]-2r^2 (d-1)(d-2)([K]+4)-r^3 (d-1)(d-2)(d-3),\nonumber\\
	\Psi_5&=&5(U_4-2U_3)+5r(d-1)(U_3-6U_2)-30r^2 (d-1)(d-2)[K]\nonumber\\
	&&-\frac{5}{2} r^3 (d-1)(d-2)(d-3)([K]+4)-r^4 (d-1)(d-2)(d-3)(d-4),\nonumber\\
	\Psi_6&=&9 U_5+15 r(d-1) (U_4-2 U_3)+10 r^2 (d-1) (d-2) (U_3-6U_2)-45 r^3 (d-1)(d-2)(d-3) [K]\nonumber\\
	&&-3 r^4 (d-1)(d-2)(d-3)(d-4) ([K]+4)-r^5 (d-1)(d-2)(d-3)(d-4)(d-5).\nonumber\\
\end{eqnarray}

\end{appendix}

\end{document}